\documentclass[letterpaper, 10pt, conference]{ieeeconf}      

\IEEEoverridecommandlockouts                              

\title{Successor Features for Transfer in Alternating Markov Games}

\author{Sunny Amatya$^{1}$, Yi Ren$^{2}$, Zhe Xu$^{2}$, and Wenlong Zhang$^{1*}$
\thanks{This work was supported by the National Science Foundation under Grant CMMI-1925403.}
\thanks{$^{1}$S. Amatya and W. Zhang are with the School of Manufacturing Systems and Networks, Arizona State University, Mesa, AZ, 85212, USA. Email:
        {\tt\small \{samatya,wenlong.zhang\}@asu.edu}}%
\thanks{$^{2}$Y. Ren, and Z. Xu are with the School for Engineering of Matter, Transport, and Energy, Arizona State University, Tempe, AZ, 85287, USA. Email:
        {\tt\small \{yiren,xzhe1\}@asu.edu}}%
        \thanks{* Address all correspondence to this author.}
}

\usepackage[utf8]{inputenc}
\usepackage{amssymb}
\usepackage{amsmath}
\usepackage{graphicx}  
\usepackage{xcolor}
\usepackage{subfigure}
\usepackage[noadjust]{cite}
\newcommand{\argmax}{\operatornamewithlimits{argmax}}
\newcommand{\argmin}{\operatornamewithlimits{argmin}}
\usepackage{color}
\usepackage{hyperref}
\usepackage{multirow}
\usepackage{adjustbox}
\usepackage{tabularx}
\usepackage{breqn}
\usepackage{booktabs}

\usepackage[linesnumbered,algoruled,boxed,lined]{algorithm2e}

\makeatletter 
\g@addto@macro{\@algocf@init}{\SetKwInOut{Parameter}{Parameters}} 
\makeatother

\iftrue 

        \newcommand{\cutparagraphup}{\vspace*{-0.17in}}
        \newcommand{\cutparagraphdown}{\vspace*{-0.03in}}

        \newcommand{\cutcaptionup}{\vspace*{-0.1in}}
        \newcommand{\cutcaptiondown}{\vspace*{-0.2in}}

        \newcommand{\cutequationup}{\vspace*{-0.07in}}
        \newcommand{\cutequationdown}{\vspace*{-0.07in}}

        \newcommand{\cuttableup}{}
        \newcommand{\cuttabledown}{}

        \newcommand{\cut}{{\vspace*{-0.02in}}}
        \newcommand{\cutmore}{{\vspace*{-0.06in}}}
        \newcommand{\negcut}{}

\else 

        \newcommand{\cutparagraphup}{
        \newcommand{\cutparagraphdown}{}

        \newcommand{\cutcaptionup}{}
        \newcommand{\cutcaptiondown}{}

        \newcommand{\cutequationup}{}
        \newcommand{\cutequationdown}{}

        \newcommand{\cuttableup}{}
        \newcommand{\cuttabledown}{}

        \newcommand{\cut}{}
        \newcommand{\cutmore}{}
        \newcommand{\negcut}{}
\fi
\usepackage{soul}
\begin{document}

\maketitle
\thispagestyle{empty}
\pagestyle{empty}
\begin{abstract}
This paper explores successor features for knowledge transfer in zero-sum, complete-information, and turn-based games. Prior research in single-agent systems has shown that successor features can provide a ``jump start" for agents when facing new tasks with varying reward structures. However, knowledge transfer in games typically relies on value and equilibrium transfers, which heavily depends on the similarity between tasks. This reliance can lead to failures when the tasks differ significantly. To address this issue, this paper presents an application of successor features to games and presents a novel algorithm called Game Generalized Policy Improvement (GGPI), designed to address Markov games in multi-agent reinforcement learning. The proposed algorithm enables the transfer of learning values and policies across games. An upper bound of the errors for transfer is derived as a function the similarity of the task. Through experiments with a turn-based pursuer-evader game, we demonstrate that the GGPI algorithm can generate high-reward interactions and one-shot policy transfer. When further tested in a wider set of initial conditions, the GGPI algorithm achieves higher success rates with improved path efficiency compared to those of the baseline algorithms.


\end{abstract}

\section{Introduction}

Multi-agent reinforcement learning (MARL) is a powerful framework for autonomous robots to coordinate in dynamic environments \cite{canese2021multi}. From multi-robot collaboration to competitive adversarial tasks, MARL has demonstrated emergent behaviors that scale to complex tasks like multi agent hide-and-seek and traffic signal control  \cite{baker2019emergent, ma2024efficient}. However, despite its success, MARL faces fundamental challenges in transferability—agents often require extensive retraining to adapt the learned policies to new tasks or agents \cite{da2019survey, amatya2022shall}. This limitation significantly hinders real-world deployment, where adaptability and efficiency are crucial.

Existing transfer learning approaches, such as curriculum learning and source task selection, require training on multiple tasks before achieving effective policy transfer \cite{sinapov2015learning, braylan2016object}. While these methods improve adaptation, they are computationally expensive and lack flexibility for robotic applications requiring humans to train in all possible setting. Other approaches focus on adapting to opponent behaviors \cite{ barrett2015cooperating}, but they are inherently reactive and fail to generalize to rapidly changing interactions. Additionally, many MARL transfer methods do not provide convergence guarantees, making them less applicable for mission-critical tasks \cite{littman2001friend}.
To address these challenges, we propose a structured transfer learning approach in turn-based Markov Games, which provide a principled framework for robust policy transfer. Our method integrates Successor Features (SF), a novel Game based Generalized Policy Improvement (GPI) with MARL to ensure stable adaptation and efficient value transfer~\cite{barreto2017successor}. Unlike prior SF-based methods \cite{liu2022efficient}, our approach leverages the alternating structure of Markov Games, allowing agents to anticipate and optimize against their opponent’s strategy. We demonstrate our approach in a minmax setting. This formulation enables transfer in adversarial settings, accounting for the worst-case opponent behavior and with  theoretical convergence guarantees, making it particularly well-suited for multi-agent applications.

\textbf{Contributions}:
In this work, we introduce a novel transfer learning algorithm for MARL. Specifically, the contributions of this paper are

\begin{itemize}
    \item Developing an SF-based minmax Q-learning algorithm that integrates novel Game based GPI for policy transfer in turn-based Markov Games.
    \item Performing convergence analysis, ensuring stable learning in competitive multi-agent settings.
    \item Empirical evaluation of our method in a two-player, minmax, turn-based pursuer-evader game, comparing it with ablations and baseline algorithms to demonstrate value transfer and efficiency of the new policy.
\end{itemize}

 The remainder of this paper is organized as follows: In Section \ref{sec:related}, related works are reviewed. The preliminaries required for is highlighted in Section \ref{sec:preliminaries}. The proposed method is described in Section \ref{sec:our_method}. Experimental results are given in Section \ref{sec:experiment} and Section \ref{sec:conclusion} concludes this paper.
\section{Related work}
\label{sec:related}

\noindent\textbf{Transfer in Games.}
One idea to transfer learning in games is to \textit{reuse past experiences}, where relevant features from previous games are leveraged to improve performance in new ones. Examples include General Game Playing (GGP) \cite{hinrichs2011transfer} and Case-Based Reinforcement Learning (RL) \cite{sharma2007transfer}, where agents build knowledge from past game play. However, these similarity-based methods require a substantial amount of experience to construct a useful base case, and ensuring convergence remains an open challenge.
A more structured approach is \textit{value function transfer}, which uses prior knowledge or shared representations to facilitate adaptation across tasks. An example is the Universal Value Function (UVF) \cite{cheng2017multi} and value function transfer in multi-agent settings \cite{liu2019value}. However, these methods struggle when reward structures or agent interaction dynamics vary significantly across games, limiting their generalizability \cite{chen2021shall}. Another policy transfer method is \textit{equilibrium transfer}, which accelerates learning by reusing policies from previous games in new ones \cite{hu2014accelerating}. This approach relies on identifying similarities in the goals of the games, allowing the agent to adapt strategies more efficiently. However, its effectiveness is limited when the goals or reward structures differ significantly, making one-shot policy transfer impractical in such cases. In this paper, we develop a framework for value and policy transfer in games with convergence guarantees, using similarity-based methods while avoiding the burden of building a library of experiences.

\noindent\textbf{Successor Features for Transfer in Reinforcement Learning.}
In transfer learning, SFs are used to model each task by defining the reward function \cite{barreto2017successor}. The ability to transfer knowledge across tasks stems from a generalization of two core RL operations: policy evaluation and policy improvement. These generalized operations, known as ``generalized policy updates", extend the standard point-based methods to set-based approaches, enabling the reuse of task solutions. When the reward function of a task can be approximated as a linear combination of reward functions for other tasks, RL can be simplified to a linear regression problem, solvable with a fraction of the data as compared to baseline Q-learning~\cite{barreto2020fast}. Additionally, the Generalized Policy Improvement (GPI) algorithm allows agents to generate policies from reference tasks by selecting the best actions accordingly.
Building on these generalized policy updates, SFs have been applied in various RL settings, including risk-sensitive environments \cite{gimelfarb2021risk} and dissimilar tasks \cite{abdolshah2021new}. While earlier work focused on agents with a single goal but different preferences, this approach has been extended to novel and unseen goals \cite{barreto2017successor}. Recent studies have explored SF-based transfer in MARL, showing that training with SFs accelerates learning by leveraging previously acquired knowledge \cite{liu2022efficient}. However, these studies have mainly focused on cooperative agents where all agents focus on single cumulative reward. In this paper, we extend the use of SFs to a new setting—transfer learning in games where two agents compete in a turn-based, zero-sum Markov game.

\section{Preliminaries and Problem Formulation}
\label{sec:preliminaries}
In this section, we give a brief background of  SFs, and two-player zero-sum Markov games. 
\subsection{Successor Features}
In Generalized Policy Evaluation (GPE), for a policy $\pi$, it takes the task $r$ as input and outputs the value function $Q^\pi_r$. 
The features for a given state and action are $\phi: S \times A \rightarrow \mathbb{R}^d$. For any feature weight, $\textbf{w} \in \mathbb{R}^d$, a task is defined using the reward function as
\begin{equation} \label{eq:onestep_reward}
    r_w (s,a) = \phi (s,a) ^T \textbf{w}
\end{equation}
where $\cdot ^T$ denotes the transpose of the vector.
Following the definition in \cite{barreto2017successor}, the SFs of a policy $\pi$ is defined as:
\begin{equation}
    \psi^\pi(s,a)^T =  \mathbb{E}^\pi [
\sum_{t=0}^{T} \gamma ^t \phi (s_{t+1}, a_{t+1})| s_t = s, \\
a_t = a
]
\end{equation}
Hence, $\psi^ {\pi} (s,a)$ gives the expected 
 discounted sum of $\phi$ when following the policy $\pi$ starting from $(s,a)$. Thus $\psi^\pi$ can be seen as a $d$-dimensional value function in which the features $\phi(s,a)$ play role in reward functions. 

Given $\psi^\pi$, one can quickly evaluate $\pi$ on a task $r_w$ by computing

\begin{equation} \label{eqn:q_mdp}
\begin{split}
\psi^\pi(s,a)^T \mathbf{w} = \mathbb{E}^{\pi} \bigg[ \sum_{t=0}^{\infty} \gamma^t 
\phi(s_{t+1}, a_{t+1})^T \mathbf{w} \,\bigg|\, s_t = s, a_t = a \bigg] \\
= \mathbb{E}^{\pi} \bigg[ \sum_{t=0}^{\infty} \gamma^t r (s_{t+1}, a_{t+1}) \,\bigg|\, s_t = s, a_t = a \bigg] 
= Q^\pi_r(s,a)
\end{split}
\end{equation}

As a consequence, SFs satisfy the Bellman equation, which means they can be computed using standard RL methods like temporal difference learning. 
\subsection{Two-player Zero-sum Markov Game}
In this paper, we consider a two-player, minmax Markov game, where two agents (ego and other) alternately take actions to maximize or minimize the cumulative discounted reward, respectively.  In alternating Markov games, since the agents take turns to make decisions, the other agent has the advantage of observing the ego's action before responding while the ego agent has the first mover's advantage. 
Following the definition in  \cite{szepesvari1996generalized}, in an alternating Markov game, the state-space is $S$,  the action-space of the ego is $A$, and the action-space of the other is $B$. The ego selects an action $a\in A$ at the current state, and the other can observe the ego’s action $a$, and selects its decision $b \in B$. Then, the system jumps to the next states with probability $P(s'|s, a,b)$, and the transition incurs with reward $r(s, a, b)$. For convenience, we consider a deterministic reward function and simply write $r(s_t, a_t, b_t)=r_t, t \in \{0,1,...\}.$
The ego’s stationary deterministic policy, $\pi: S \to A$, maps a state $s \in S$ to an action $\pi(s) \in A$, while the other’s stationary deterministic policy, $\mu: S \times A \to B$, maps a state $s \in S$ and the ego’s action $a \in A$ to an action $\mu(s, a) \in B$.
 It has been shown that there exists an optimal stationary deterministic policy for both the ego and other agents \cite{littman1994markov}.


The objective of the Markov game is to determine the ego’s optimal policy, denoted as $\pi^*$, and the optimal adversarial policy, denoted as $\mu^*$:
\begin{equation}
 (\pi^*,\mu^*):= arg\max_{\pi \in \Theta} \min_{\mu \in \Omega}  \mathbb{E}\begin{bmatrix}
\sum_{t = 0}^{\infty} \gamma ^t r_t|\pi, 
\mu
\end{bmatrix} 
\end{equation}  \label{eqn:markov_game_policy}

\noindent where $\gamma \in [0,1)$  is the discount factor, $\Theta$ and $\Omega$ are the sets of all admissible deterministic policies of the ego and the other agent, respectively. $(s_0,a_0,b_0,s_1,a_1,b_1,...)$ is a state-action trajectory generated under policies $\pi,\mu$, and $\mathbb {E} [\sum_{t=0}^{\infty} \gamma ^j r_t|\pi,\mu] $ is an expectation conditioned on the policies $\pi$ and $\mu$, which looks similar to \eqref{eqn:q_mdp}. Thus, we propose to decompose the Q-values into the SF values, as highlighted in the Sec. \ref{subsec:GGPIwSF}. 

Some fundamental tools used in Markov decision processes, such as the value function and Bellman equation, can be also applied to the two-player Markov game \cite{bertsekas2012dynamic}. In particular, the optimal Q-function is defined as

\begin{multline}
 Q^*(s,a,b):= \max_{\pi \in \Theta} \min_{\mu \in \Omega}  \mathbb{E}\begin{bmatrix}
\sum_{t = 0}^{\infty} \gamma ^t r_t|s_t= s, a_t = a, \\ b_t= b, \pi, \mu
\end{bmatrix} 
\end{multline}
which satisfies the optimal Q-Bellman equation 
\begin{multline}
  Q^*(s,a,b):=  r(s,a,b)+ \\ \gamma \sum_{s' \in S} P(s'|s,a,b) \max_{a'\in A} \min_{b' \in B}  Q^*(s,a',b') 
\end{multline}

\noindent The ego’s stationary optimal policy is given by 
\begin{align} \label{eqn:ego_policy}
\pi^*(s)=arg \max_{a \in A} \min_{ b\in B} Q^*(s, a, b),
\end{align}
and the other’s stationary optimal policy is
\begin{align}
\mu^*(s,a)=arg\min_{b\in B} Q^*(s,a,b).
\end{align}
Using the Bellman equation, the Q-value iteration (Q-VI) is the recursion 
\begin{align}
Q_{t+1}(s,a,b) = (FQ_t)(s, a,b), \forall (s,a,b)\in S\times A\times B, 
\end{align}
where $F$ is the contraction operator. It is known that the Q-VI converges to $Q^{*}$ \cite{zhu2020online}.
Let $\Pi = \{\pi, \mu\}$ for the sake of ease moving forward. 
\section{Transfer via Successor Features}
\label{sec:our_method}
We now consider scenarios where the source and target tasks have distinct reward functions. The goal of transfer in MARL is to learn an optimal policy for the target task by utilizing knowledge gained from the source task. We first define the set of source task models using \eqref{eq:onestep_reward} as follows
\begin{equation}\label{eqn:transfer_problem}
\mathcal{M}^\phi = \{M(S,A,B,p,r,\gamma)|r(s,a,b)= \phi(s,a,b)^T\textbf{w}\}.
\end{equation} 
At the time when the agent wants to learn a target task, it already has a collection of good policies for a set of source tasks in  $\mathcal{M} =  \{M_1, M_2, ..., M_n\}$ with  $M_i \in \mathcal{M}^\phi$. Hence for a new task $M_{n+1}$, we are able to get a value function utilizing $\mathbf{w_{n+1}}$. This value generation is before we have carried out any policy improvement in the target task. Let the initial policy that the agent computes be $\Pi'$. At this point, we can observe one-shot transfer. In our experiments, we will be able to show this as well. If we continue learning in the new task, the resulting policy $\Pi$ should be such that  $Q^\Pi(s,a, b) \geq Q^{\Pi'}(s,a,b)$, $\forall (s,a,b) \in S \times A \times B$. 




With this setup, we can now describe our approach for solving the transfer problem outlined in \eqref{eqn:transfer_problem}. Our solution is presented in two stages: we first generalize policy improvement for games, and then demonstrate how SFs can be effectively utilized to implement this generalized policy improvement.

\subsection{Game Generalized Policy Improvement}
In this section, we highlight our proposed policy improvement algorithm for games, and explain how the new policy is computed based on the $Q$-value of a set of policies. We show that this extension can be a natural greedy policy for both agents over the available aforementioned $Q$-value functions.  

Generally for a single-task multi-agent game, the policy iteration algorithm initially chooses an arbitrary pair of strategies and repeatedly selects a better pair of strategies until a local optimal pair of strategies is found. This is further highlighted in Sec. \ref{eqn:markov_game_policy} where both agents try to find their optimal policy by choosing greedy policies in the available $Q$ values~\cite{howard1960dynamic}. 
For multiple tasks, we propose Game Generalized Policy Improvement (GGPI) as follows.

\textbf{Theorem 1. (Game Generalized Policy Improvement)}.
Let $\Pi_1, \Pi_2, ..., \Pi_n $ be $n$ policies and let $\tilde Q ^{\Pi_1}, \tilde Q^{\Pi_2}, ..., \tilde Q ^{\Pi_n}$ be approximations of their respective $Q$ functions such that
\begin{dmath} \label{eqn:bound}
|Q ^{\Pi_i}(s,a, b)- \tilde Q ^{\Pi_i}(s,a, b)| \leq \epsilon \\ \forall s \in S , a \in A, b \in B \text{ and } i \in \{1,2,... n\}. 
\end{dmath}
Define $
\pi_{GGPI}(s) \in \argmax_a \min_b \min_i \tilde Q^{\Pi_i} (s,a, b)$, one can show
\begin{equation}\label{eqn:bound_greater}
Q^\Pi(s,a,b) \geq \min_i Q^{\Pi_i} (s,a,b) - \frac{2 \epsilon}{1- \gamma}.
\end{equation}


The proof is provided in Supplementary Materials I\footnote{Supplementary materials can be found at \url{https://home.riselab.info/downloads/IROS_2025_SuppMaterials.pdf}}. Our theorem accounts for cases where the value functions of policies are not computed exactly, either due to the use of function approximation or because an exact algorithm has not been fully executed. This approximation error is captured in \eqref{eqn:bound}, with $\epsilon$ which appears as a penalty term in the lower bound \eqref{eqn:bound_greater}. As stated in \cite{barreto2017successor}, such a penalty is inherent to the use of approximation in RL and is identical to the penalty incurred in the single-policy case (Proposition 6.1 in \cite{bertsekas1996neuro}). In this context, Theorem 1 guarantees that the selected policy $\Pi$ will perform no worse than any of the policies $\Pi_1, \Pi_2, ..., \Pi_n$. This result extends naturally to multi-agent Markov games, where the policy of an agent, as defined in  \eqref{eqn:ego_policy}, is generalized to a set of policies across agents.

In this work, we analyze two key aspects: 1) how learning new tasks and the use of SFs facilitate the transfer process, and 2) the analysis of multiple policies across different initial conditions to enable effective transfer using SF.

\subsection{Game Generalized Policy Improvement with Successor Feature} \label{subsec:GGPIwSF}

We start this section by extending our notation slightly to make it easier to refer to the quantities involved in transfer learning. Let $M_i$ be a task in $\mathcal{M}^\phi$ defined by $\mathbf{w_i}$  $\in \mathbb{R}^d$. We use $\Pi_i^*$ to refer to an optimal policy of $M_i$ and use $Q^{\Pi_i^*}_i$ to refer to its value function. The value function of $\Pi_i^*$ when executed in $M_j \in\mathcal{M}^\phi$ will be denoted by $Q^{\Pi_i^*}_j$.

Suppose now that an agent has computed optimal policies for the tasks $M_1, M_2, \ldots, M_n \in\mathcal{M}^\phi$. When presented with a new task $M_{n+1}$, the agent computes $\{Q^{\Pi_1^*}_{n+1}, Q^{\Pi_2^*}_{n+1}, \ldots, Q^{\Pi_n^*}_{n+1}\}$, the evaluation of each $\Pi_i^*$ under the new reward function induced by $\mathbf{w_{n+1}}$. 



Suppose that we have learned the functions $Q^{\Pi_i^*}_i$ using the representation shown in \eqref{eqn:q_mdp}. Now, if the reward changes to $r_{n+1}(s, a, b, s') = \phi(s, a, b, s')^T \mathbf{w_{n+1}}$, as long as we have $\mathbf{w_{n+1}}$, we can compute the new value function of $\Pi_i$ by simply making $Q^{\Pi_i^*}_{n+1}(s, a, b) = \psi^{\Pi^*_i}(s, a, b)^T\mathbf{w_{n+1}}$. This significantly simplifies the computation of all $Q^{\pi_i^*}_{n+1}$. 


Once the functions $Q^{\pi_i^*}_{n+1}$ have been computed, we can apply GPI to derive a policy $\Pi$ whose performance on $M_{n+1}$ is no worse than the performance of $\Pi_1^*, \Pi_2^*, \ldots, \Pi_n^*$ on the same task. A question that arises in this case is whether we can provide stronger guarantees on the performance of $\Pi$ by exploiting the structure shared by the tasks in $\mathcal{M}^\phi$. The following Lemma answers this question in the affirmative.

\textbf{Lemma 1.} Let $\delta_{ij} = \max_{s, a, b} \lvert r_i(s, a, b) - r_j(s, a, b) \rvert$ and let $\Pi$ be an arbitrary policy. 
Then,
\begin{equation} \label{eqn:bound_upper}
\lvert Q^{\Pi}_i(s, a, b) - Q^{\Pi}_j(s, a, b) \rvert \leq \frac{\delta_{ij}} {1 - \gamma}. 
\end{equation}

The detailed proofs along with additional information can be found in the Supplementary Materials. This lemma provides an upper bound of the differences in action values of the transfer policy in \eqref{eqn:bound_upper} and the optimal policies of the target task $M_j$. As stated in \cite{barreto2017successor}, this provides the guidance for application of the algorithm for storing and removing the SFs. It further captures the role of the reward $r$ which can be further broken down into feature $\phi$ and the weight $\mathbf{w}$. Hence, one can show $\delta_{ij} = \max_{s, a, b}\lvert\phi\rvert \lvert\mathbf{w_{r,i} -w_{r,j}} \rvert$, where the weights characterize the difference in the task objective. 

\noindent\textbf{Algorithm and practical use.}
Algorithm \ref{algo:SFminmaxQ} outlines our solution using turn-based minmax Q-learning as the baseline RL method. In this approach, both the ego agent and the other agent perform a one-step lookahead into their partner's actions, choosing moves to either maximize or minimize rewards based on the opponent's choices. Our algorithm highlights the use of GGPI by the ego and other agents, as shown in lines 7 and 18, respectively. This allows the policies $\Pi_i$ to be used to solve each task $M_i$. 

\section{Experiments}
\label{sec:experiment}
In this section, we first introduce the testing setup which includes the environment definitions, experimental conditions, evaluation metrics and the baselines. Then we show the comparison of our algorithm with the baseline algorithms. 
\subsection{Simulation Setup}\label{subsec:simulation_setup}
\begin{algorithm}[t]
\SetAlgoLined
\SetKwInOut{Input}{input}
\SetKwInOut{Output}{output}
\Input{discount factor $\gamma$, task index $k$, \\learning rate $\alpha$}
\Output{Successor features  $\psi_i^{\pi_k}$}
    Initialize $Q_0 \in \mathbb{R}^{|S\times A\times B|}$\\
    \For{ns steps}{
        \For {ego agent}
        {       
            \uIf{Bernoulli($\epsilon)=1$}{$a \leftarrow$ Uniform ($A$)}

             \Else { $a_t, b_t= \argmax_b \argmin_a \min_i \Psi^{\Pi_i} (s,a,b)^T \mathbf{w_i}$}
            Execute actions $a_t$ and observe feature and reward $\phi_t$, $r_t$ and new state $s_{t}'$\\
            \uIf{new task}{
                $ a' b' =  \argmax_b \argmin_a \Psi^{\pi_{n+1}} \mathbf{w_{n+1}}$ \\
                $\Psi^{\pi_{n+1}}_{t+1}(s_t, a_t, b_t) = \Psi^{\pi_{n+1}}_{t}(s_t, a_t, b_t)+ \alpha \begin{Bmatrix}{\phi_t} +
               \Psi_t(s', a', b') - \Psi_t(s_t, a_t, b_t)\end{Bmatrix}$ 
                }
            \uIf {$s_{t}'$ is not terminal}{
                  \For {other agent}{

                  \uIf{Bernoulli($\epsilon)=1$}{$a \leftarrow$ Uniform ($A$)}

                     \Else { $a_t, b_t= \argmax_b \argmin_a \min_i \Psi^{\Pi_i}(s_t',a,b)^T \mathbf{w_i}$}
                    Execute actions $a_t$ and observe feature and reward $\phi_t$, $r_t$ and new state $s_{t+1}'$\\
                    \uIf{new task}{
                        $ a' b' =  \argmax_b \argmin_a \Psi^{\pi_{n+1}} \mathbf{w_{n+1}}$ \\
                        $\Psi^{\pi_{n+1}}_{t+1}(s_t', a_t, b_t) = \Psi^{\pi_{n+1}}_{t}(s_t', a_t, b_t)+ \alpha \begin{Bmatrix}{\phi_t} +
                       \Psi_t(s_{t_1}', a', b') - \Psi_t(s_t, a_t, b_t)\end{Bmatrix}$ 
                        }
                  }
                }

            }
        }
\caption{ SFminmax: Turn-based Minmax Q learning with SF and GGPI}
\label{algo:SFminmaxQ}
\end{algorithm}
\subsubsection{\textbf{Environment definition}}\label{subsubsec:environment}
The experiments are conducted using a pursuer-evader game on a $5 \times 5$ grid, as shown in Fig.~\ref{fig:scenarios}. The evader/ego agent is represented by a robber, the pursuer/other agent by a police officer, and the goal position by an exit door. The action space includes four possible actions, $A= \{\text{up}, \text{down}, \text{left}, \text{right}\}$. The key features include the Manhattan distance from the evader to the three possible goals (exits), $\{d (x_e-g_1), d (x_e-g_2), d (x_e-g_3)\}$,  and the Manhattan distance between the agents, $d(x_e-x_o)$ and a terminal reward that the evader receives upon reaching the specific goal, $\{r_t(g_1), r_t(g_2), r_t(g_3)$\}.


To normalize the reward, we divide the two distance-based features $d$ by the maximum possible distances between agents and the distance of evader from the goal. 
If the evader successfully reaches the goal, it earns a token with a value of 0.7, which is chosen based on parameter tuning. The feature weights determine the reward associated with each feature, and the weights are reset each time a new task begins.

An example of the feature weights used for training and testing the environment is: $[0.7,-1.3, 0.7, 0, 0, 0, 0 ]$. The corresponding feature vector is defined as: $\phi= [d(x_e-x_o), d (x_e-g_1), r_t(g_1), d (x_e-g_2), r_t(g_2), d (x_e-g_3), r_t(g_3)]$. From the ego agent’s perspective, it receives a positive reward for maintaining a greater distance from other agents, incurs a high loss when far from its goal, and obtains a terminal reward upon reaching its goal. The reward for Task 1 is computed by multiplying the feature vector, $\phi$, with the weight vector, $\mathbf{w_1}$ following \eqref{eq:onestep_reward}. The weights for Tasks 2 and 3 are provided in Table \ref{table:weight}.

We run each game (task) for a designated number of iterations or episodes. In the baseline scenario, the agents reset at the start of each new task. However, in our approach, the agents carry over the SF table from the previous run. Initially, both agents are placed at equal distances from the goal. For a more quantitative analysis, the agents' starting positions are randomized after each episode, i.e., when the evader reaches the goal or the pursuer catches the evader. Further details of the experiment is explained in Sec. \ref{subsec:quantitative}.

\begin{table}
\caption{Table of tasks and the weights associated with the tasks}
\begin{center}
\begin{tabular}{ |c| c |}
\hline
$M_i$ & weight ($\mathbf{w_i}$) \\
\hline
 Task 1 & [0.7,-1.3, 0.7, 0, 0, 0, 0 ] \\ 
 Task 2 & [0.7, 0, 0,-1.3, 0.7, 0, 0 ] \\  
 Task 3 & [0.7, 0, 0, 0, 0,-1.3, 0.7]   \\
 \hline
\end{tabular}
\label{table:weight}
\end{center}
\vspace{-0.1in}
\end{table}

\subsubsection{\textbf{Compared Method and Evaluation Metrics}}\label{subsubsec:metrics_game}

The performance of the ego agent is measured by the cumulative reward during training, with MinMaxQ-learning \cite{lee2023finite} as the baseline. We further assess the policy generated across different tasks, where both (ego and other) agents use the same learning algorithm. 

Additionally, we compare our method to other transfer learning algorithms, including the Probabilistic Policy Reuse (PRQL) algorithm, which facilitates policy transfer between tasks \cite{fernandez2010probabilistic}. Here, PRQL has been adapted for multi-agent scenarios. Furthermore, we carry an ablation test with the $\epsilon$ reset version of SFminmaxQL as SF-r where at the beginning of each task we reset $\epsilon$ from Algorithm \ref{algo:SFminmaxQ} line 4 to promote more exploration. For the sake of brevity, we will refer to MinMaxQ-learning as ``Minmax" and the original SFminmaxQL as ``SFminmax".

The aformentioned algorithms are further analyzed for robustness under different initial conditions. We evaluate the discrepancy in values, denoted as $|V - V'|$, across all possible initial positions for various tasks and episode lengths. Here, $V'$ refers to the value at 2,000 episodes for the minmax algorithm, with a lower value indicating better performance.

We also report the success rate (SR), defined as the percentage of wins (pwin/ewin) and ties (tie) in each game. An episode is considered successful if the pursuer catches the evader or if the evader escapes within a 30-step limit. Otherwise, the episode is recorded as a tie. In addition, we report Success Weighted by Path Length (SPL) \cite{anderson2018evaluation}, which accounts for the efficiency of the navigation process. SPL$= \frac{1}{N} \sum_i^N S_i \frac{l_i}{\max(p_i, l_i)}$ where $p_i$ is the path taken by the baseline agent,000 episodes), $l_i$ is the path taken by trained agent at given episode, $S_i$ is the binary indicator of success for episode $i$, and $N$ is the total number of episodes. The baseline algorithm will reach a value of 1 by the end of the training  while for the other trained agents, higher SPL scores indicates shorter path.


%


\subsection{Reward Transfer: SFminmax vs. Baseline}

\noindent\textbf{Experimental setup.}
The baseline used is the standard Minmax algorithm. Using the evaluation metrics discussed in Sec.~\ref{subsubsec:metrics_game}, we compare the baseline with the proposed SFminmax algorithm. In the first case study, both agents start at equal distances from the goal. The goal position remains fixed for the first 30,000 iterations, after which it changes. At this point, the baseline algorithms reset, while the proposed SFminmax algorithm continues using the SF table. We adopt a discount factor of $\gamma = 0.9$ and a learning rate of $\alpha = 0.5$. In this experiment, we focus on the cumulative reward during the training phase, where the reward represents the ego agent’s reward minus that of the other agent.

\begin{figure}[t]
\centering
    \vspace{-0.05in}
    \includegraphics[width=1\linewidth]{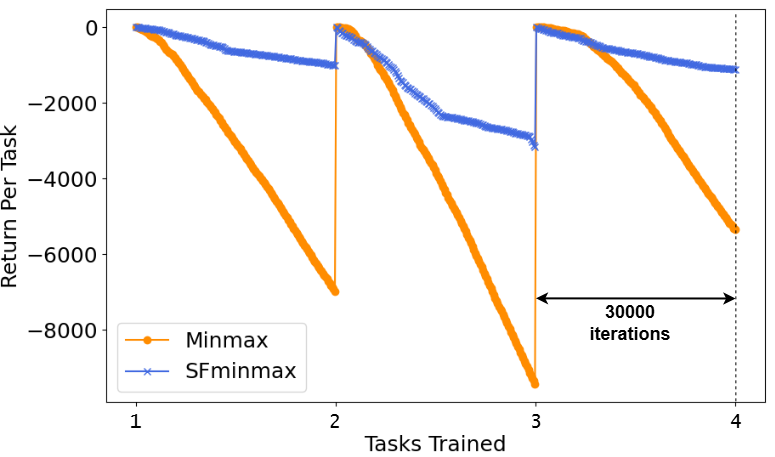}
    \caption{Cumulative return per task in the Pursuer-Evader game. SFminmax reward uses the feature from the previous task at the beginning to each task.}
    \label{fig:training_reward}
\end{figure}
\noindent\textbf{Observation.}
Our results demonstrate that SFminmax successfully transfers knowledge from one task to another, resulting in higher cumulative return for the reward during training compared to the baseline algorithm as seen in Fig. \ref{fig:training_reward}.

\subsection{Test Policy Transfer: SFminmax vs. Baseline}
\noindent\textbf{Experimental setup.}
The experimental setup mirrors the one used in the previous case study. However, instead of evaluating the value function, we focus on the agents' policies. As outlined in the previous section, testing is conducted every 10,000 iterations and at the start of each new task. Here, we examine the agents' policies at specific iterations to assess whether the transfer of knowledge is successful.

\begin{figure*}[h]
    \centering
    \vspace{-0.05in}
    \includegraphics[width=1\linewidth]{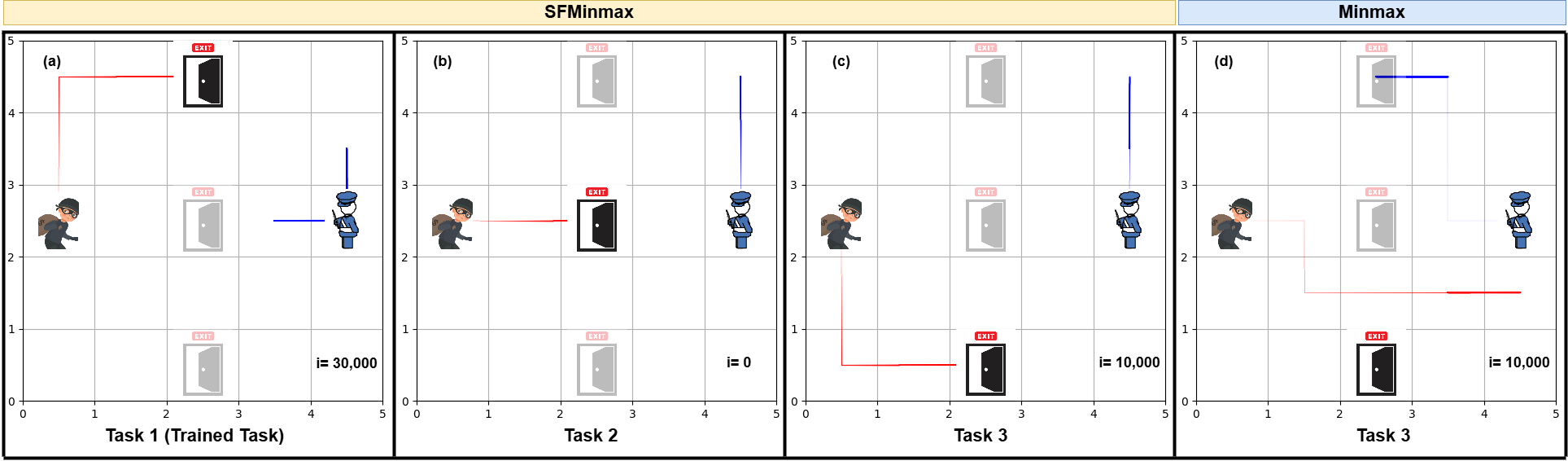}
    \caption {Trajectories taken by the evader (robber) and pursuer (police) for three tasks determined by the goal position (exits).  The trajectories for case study (a) at the end of 30,000 iterations have the evader reach the goal. (b) shows the one-shot transfer of the policy when the goal position changes. (c) The trajectory takes 10000 more interaction to converge in Task 2. (d) The baseline does not converge with the same number of episodes.}
     \label{fig:scenarios}
\end{figure*}

\noindent\textbf{Observation.}
SFminmax can transfer knowledge across tasks within the policy space, as seen in Fig.~\ref{fig:scenarios}. The baseline Minmax algorithm struggles to converge to a model when the number of iterations is limited. In contrast, the proposed algorithm provides a jump-start for new tasks, enabling the generation of converging policies. When transferring from Task 1 to Task 2, we observe near-instantaneous results, with policies converging after 0 iterations as seen in  Fig.~\ref{fig:scenarios}(b). For Task 3, however, due to the different end-goal positions, policy convergence occurs within 10,000 iterations as seen in Fig.~\ref{fig:scenarios}(c), whereas the baseline algorithm fails to achieve similar performance within 10,000 iterations as shown in Fig.~\ref{fig:scenarios}(d).

\begin{table*}[h]
\caption{Navigation performance comparison in different test tasks}
\setlength\tabcolsep{5.2pt}
\begin{tabular}{|ccccccccccccc|}
\hline
\multicolumn{1}{|c|}{}     & \multicolumn{4}{c|}{$|V-V'|$}                                                  & \multicolumn{4}{c|}{SR \% {[}ewin, tie{]}}                                                                           & \multicolumn{4}{c|}{SPL}                      \\ \hline
\multicolumn{1}{|c|}{Eps}  & Minmax & SFminmax       & SF-r           & \multicolumn{1}{c|}{PRQL}           & Minmax           & SFminmax           & Sf-r                                       & \multicolumn{1}{c|}{PRQL}             & Minmax & SFminmax      & Sf-r          & PRQL \\ \hline
\multicolumn{13}{|c|}{\textbf{Task 2}}                                                                                                                                                                                                                                                   \\ \hline
\multicolumn{1}{|c|}{0}    & 0.535  & \textbf{0.07}  & \textbf{0.345} & \multicolumn{1}{c|}{1.03}           & {[}55.5, 33.3{]} & \textbf{{[}100, 0{]}}    & {[}66.7, 33.3{]}                           & \multicolumn{1}{c|}{{[}22.2, 33.3{]}} & 0.46   & \textbf{1.06} & 0.70          & 0.2  \\
\multicolumn{1}{|c|}{1000} & 0.01   & 0.03  & 0.452 & \multicolumn{1}{c|}{0.11}           & {[}100, 0{]}     & {[}100, 0{]}       & {[}44.4,55.6{]}                            & \multicolumn{1}{c|}{{[}77.8, 0.0{]}}  & 0.80   & 1.06          & 0.43          & 0.89 \\
\multicolumn{1}{|c|}{2000} & NA     & 0.128          & 0.345          & \multicolumn{1}{c|}{0.07}           & {[}88.8, 0{]}    & {[}88.8, 0.0{]}    & {[}55.5, 44.4{]}                           & \multicolumn{1}{c|}{{[}77.8, 0.0{]}}  & 1.0    & 1.09          & 0.62          & 0.83 \\ \hline
\multicolumn{13}{|c|}{\textbf{Task 3}}                                                                                                                                                                                                                                                   \\ \hline
\multicolumn{1}{|c|}{0}    & 0.900  & \textbf{0.141} & 0.331 & \multicolumn{1}{c|}{0.612}          & {[}22.2. 11.1{]} & \textbf{{[}88.9, 0.0{]}} & \textbf{{[}88.8, 0.0{]}}                         & \multicolumn{1}{c|}{{[}44.44, 22{]}}  & 0.43   & \textbf{1.29} & \textbf{1.15} & 0.26 \\
\multicolumn{1}{|c|}{1000} & 0.176  & 0.193 & 0.210          & \multicolumn{1}{c|}{\textbf{0.15}}  & {[}100, 0{]}     & {[}100, 0.0{]}     & {[}100.0, 0{]}                             & \multicolumn{1}{c|}{{[}77.78, 0.0{]}} & 1.0    & 1.15 & 0.74          & 1.15 \\
\multicolumn{1}{|c|}{2000} & NA     & 0.458          & \textbf{0.158 }         & \multicolumn{1}{c|}{\textbf{0.17}}  & {[}88.8, 0{]}    & {[}77.8, 11.1{]}   & {[}88.8, 0.0{]}                            & \multicolumn{1}{c|}{{[}55.6, 0.0{]}}  & 1.0    & 1.05 & 0.97          & 1.05 \\ \hline
\multicolumn{13}{|c|}{\textbf{Task 4}}                                                                                                                                                                                                                                                   \\ \hline
\multicolumn{1}{|c|}{0}    & 0.87   & \textbf{0.621} & \textbf{0.355} & \multicolumn{1}{c|}{1.307}          & {[}33.3, 44.4{]} & {[}44.4, 0.0{]}    & {[}88.8, 0.0{]}                            & \multicolumn{1}{c|}{{[}11.1, 22.2{]}} & 0.24   & \textbf{1.39} & \textbf{1.68} & 0.5  \\
\multicolumn{1}{|c|}{1000} & 0.128  & 0.319          & 0.505          & \multicolumn{1}{c|}{0.197}          & {[}100, 0{]}     & {[}66.7. 11.1{]}   & {[}66.6, 0.0{]}                            & \multicolumn{1}{c|}{{[}55.6, 0{]}}    & 0.74   & 1.19 & 1.09 & 0.91 \\
\multicolumn{1}{|c|}{2000} & NA     & 0.448          & 0.44           & \multicolumn{1}{c|}{\textbf{0.180}} & {[}88.8, 0{]}    & {[}\textbf{66.7, 0.0}{]} & {[}\textbf{66.6, 0.0}{]} & \multicolumn{1}{c|}{{[}77.8, 0{]}}    & 1.0    & 1.09 & 1.17 & 1.12 \\ \hline
\end{tabular}
\label{tab:qualitative}
\vspace{-0.1in}
\end{table*}
\subsection{Transfer with Significant Changes in Reward} 
\label{subsec:quantitative}
\begin{figure*}[t]
\centering
    \vspace{-0.05in}
    \includegraphics[width=1\linewidth]{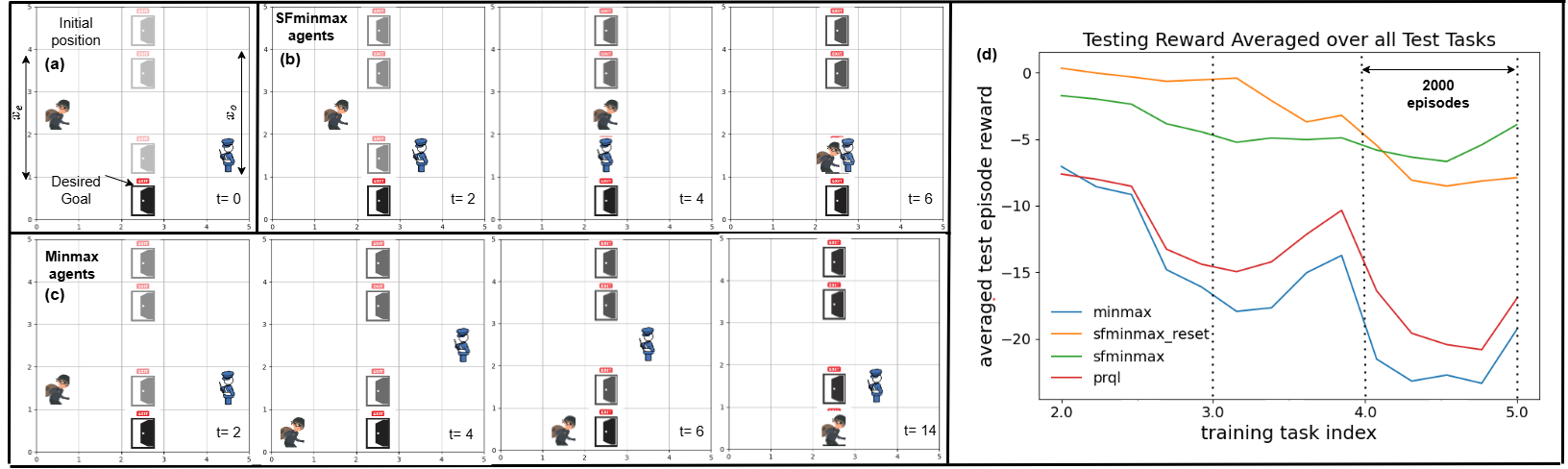}
    \caption{(a) Possible initial position and goals for testing transfer with significant reward change (b) Path taken by SFminmax agents where pursuer agent is able to intervene. (c) Path taken my Minmax agents where pursuer is not to intervene (d) Cumulative return per task in the Pursuer-Evader game. SFminmax reward uses the feature from the previous task at the beginning to each task.}
    \label{fig:testing_reward}
    \vspace{-0.05in}
\end{figure*}

\noindent\textbf{Experimental setup.}
We first pre-train the model where the initial positions, $x_e$ and $x_{o}$, are randomly sampled from $x_e \in [1,2,3]$ for the $x$-coordinate, where $o$ and $e$ stand for the pursuer/other and evader/ego in the game, respectively. The $y$-coordinate fixed at 0.  This pre-training is done on Task 1, where the goal position is set to (2,5) as shown in Fig. \ref{fig:testing_reward}(a). After pre-training, we retrain the model on new tasks with changing goal positions: Task 2 has a goal at (2, 4), Task 3 at (2, 1), and Task 4 at (2, 0).
SFs for different initial and goal conditions for single agent have been previously studied in \cite{lehnert2017advantages}. In this work, we explore how SFs with GGPI facilitates transfer to new goal positions in multi agent games. During training, we evaluate the performance at the midpoint of the game and examine the results after final convergence.
%


\noindent \textbf{Hyperparameter tuning.} 
We use a decaying $\epsilon$ for each new task, following the method proposed in \cite{lehnert2017advantages}, for all algorithms except for SFminmax. 
We conducted an ablation study with learning rates $\alpha \in [0.1, 0.2, 0.3, 0.4, 0.5]$ from Algorithm \ref{algo:SFminmaxQ} line 12 and baseline algorithms to identify the significance of priority of new to old experience in all temporal difference algorithms. We found the best results for the baseline algorithm with $\alpha_{\text{minmax}} = 0.3$ and for our algorithm with $\alpha_{\text{sf}} = 0.1$.

\noindent \textbf{Observation.}
First, we analyze the average reward of the ego agents, averaged over nine initial conditions. The rewards are sampled five times, every 500 episodes from 0-2000 episode per task. We observe that the normalized reward values for agents using SFs are higher compared to those without SF, as shown in Fig.~\ref{fig:testing_reward}(b). To further investigate the optimality of the solution, we also summarize the results in Table \ref{tab:qualitative}. The key observations are as follows.

\noindent\underline{One-shot transfer}: The one-shot transfer of both SF-based algorithms (SFminmax and SF-r) perform better compared to random restart from the other algorithms as seen by lower $|V-V'|$ at timestep 0 for SF algorithms. 

\noindent\underline{Success rates}: As episodes increase (e.g., from 0 to 2,000), some algorithms show variations in SR\%. The reduction on the tie in SR\% indicates learning for both Minmax and PRQL algorithm.
However, we observe the same win rate in all the three tasks for Minmax, indicating bias. Here, the evader agent  wins in eight out of nine scenarios, even in scenarios where the pursuer has advantage. This can be seen in Fig.~\ref{fig:testing_reward}(a). At $t= 0$, the pursuer is closer to goal and has advantage. In  Fig.~\ref{fig:testing_reward}(c) for Minmax agents, the pursuer is not able to intervene and evader agent reaches the goal at $t=14$. However, the pursuer agent is able to catch the evader agent and finish the game within $t=6$, as shown in Fig.~\ref{fig:testing_reward}(b). 

\noindent\underline{Efficiency of Policy}: Among the successful tasks we see that SFminmax consistently shows higher SPL values across tasks compared to Minmax, SF-r and PRQL. This shows that transfer in value and policy lead to shorter path length. 
\noindent\underline{Impact of Task Variability}: 
As the task keeps changing during training, a significant deviation from the original task leads to performance degradation in SFminmax, particularly when comparing $|V-V'|$ from Task 2 to Task 4 at 2000 episodes.SF-r, however, outperforms the original SFminmax as observed in value discrepancy for Task 3. Additionally, PRQL algorithms tend to perform better with more training episodes and when the test task differs more significantly from the trained task. This is evident in Task 4, where PRQL achieves lower $|V-V'|$ as well as higher SPL scores. The value discrepancy arises party differences in agent wins, as reflected in SR\%. For instance, Minmax achieves an ewin/pwin rate of 88.9\%/11.1\% in Task 4, whereas SF algorithms result in a 66.7\%/33.3\% split.

\section{Conclusion, Limitation and Discussion}
\label{sec:conclusion}
In this paper, we introduced an algorithm that leverages SFs with GGPI to enable transfer in alternating Markov games. We analyzed the convergence properties and establish a transfer bound for the novel GGPI algorithm, extending the GPI for games. We evaluated the proposed algorithm in a grid-based pursuer-evader game, demonstrating successful policy and value transfer for cases with the same initial positions. To test transfer with significant change in reward, we further carried out an experiment with different initial  and goal positions for our proposed algorithm and other baseline algorithms. Here, we showed that our proposed algorithm performs better in terms of initial value transfer, shorten the time taken to complete the game in successes, and generate higher normalized reward for all the tasks and initial positions. Finally, we highlighted the advantages of exploration when the test task significantly differs from the training task by evaluating a reset version of our transfer algorithm (SF-r) alongside PRQL.

Notably, the SFminmax and SF-r approaches, while showing significant promise, have certain limitations. The use of the GGPI algorithm requires a discrete action space and assumes a conservative strategy by modeling the opponent's best response. Additionally, it relies on the Q-values of the training and test tasks being within comparable bounds. While this approach works well when tasks have similar reward structures or importance, it becomes less effective when the nature or weighting of the tasks changes significantly. Consequently, the proposed algorithm can be extended to a continuous state environment where further tests can be carried out in similar zero-sum game tasks with convergence guarantees \cite{lowe2017multi, yu2020meta}. Similarly, SFs can be explored in turn-based extensive-form games, where a new policy improvement algorithm could be developed to both establish transfer bounds and enhance successful transfer.

\section*{ACKNOWLEDGMENT}
The authors would like to thank Mukesh Ghimire and Lei Zhang for assistance with verifying the algorithm.


\bibliography{ref}

\begin{thebibliography}{1}
\providecommand{\url}[1]{#1}
\csname url@samestyle\endcsname
\providecommand{\newblock}{\relax}
\providecommand{\bibinfo}[2]{#2}
\providecommand{\BIBentrySTDinterwordspacing}{\spaceskip=0pt\relax}
\providecommand{\BIBentryALTinterwordstretchfactor}{4}
\providecommand{\BIBentryALTinterwordspacing}{\spaceskip=\fontdimen2\font plus
\BIBentryALTinterwordstretchfactor\fontdimen3\font minus
  \fontdimen4\font\relax}
\providecommand{\BIBforeignlanguage}[2]{{%
\expandafter\ifx\csname l@#1\endcsname\relax
\typeout{** WARNING: IEEEtran.bst: No hyphenation pattern has been}%
\typeout{** loaded for the language `#1'. Using the pattern for}%
\typeout{** the default language instead.}%
\else
\language=\csname l@#1\endcsname
\fi
#2}}
\providecommand{\BIBdecl}{\relax}
\BIBdecl

\bibitem{littman1996generalized}
M.~L. Littman and C.~Szepesv{\'a}ri, ``A generalized reinforcement-learning
  model: Convergence and applications,'' in \emph{ICML}, vol.~96, 1996, pp.
  310--318.

\bibitem{barreto2017successor}
A.~Barreto, W.~Dabney, R.~Munos, J.~J. Hunt, T.~Schaul, H.~P. van Hasselt, and
  D.~Silver, ``Successor features for transfer in reinforcement learning,''
  \emph{Advances in neural information processing systems}, vol.~30, 2017.

\end{thebibliography}


\begin{thebibliography}{10}
\providecommand{\url}[1]{#1}
\csname url@samestyle\endcsname
\providecommand{\newblock}{\relax}
\providecommand{\bibinfo}[2]{#2}
\providecommand{\BIBentrySTDinterwordspacing}{\spaceskip=0pt\relax}
\providecommand{\BIBentryALTinterwordstretchfactor}{4}
\providecommand{\BIBentryALTinterwordspacing}{\spaceskip=\fontdimen2\font plus
\BIBentryALTinterwordstretchfactor\fontdimen3\font minus
  \fontdimen4\font\relax}
\providecommand{\BIBforeignlanguage}[2]{{%
\expandafter\ifx\csname l@#1\endcsname\relax
\typeout{** WARNING: IEEEtran.bst: No hyphenation pattern has been}%
\typeout{** loaded for the language `#1'. Using the pattern for}%
\typeout{** the default language instead.}%
\else
\language=\csname l@#1\endcsname
\fi
#2}}
\providecommand{\BIBdecl}{\relax}
\BIBdecl

\bibitem{canese2021multi}
L.~Canese, G.~C. Cardarilli, L.~Di~Nunzio, R.~Fazzolari, D.~Giardino, M.~Re,
  and S.~Span{\`o}, ``Multi-agent reinforcement learning: A review of
  challenges and applications,'' \emph{Applied Sciences}, vol.~11, no.~11, p.
  4948, 2021.

\bibitem{baker2019emergent}
B.~Baker, I.~Kanitscheider, T.~Markov, Y.~Wu, G.~Powell, B.~McGrew, and
  I.~Mordatch, ``Emergent tool use from multi-agent autocurricula,''
  \emph{arXiv preprint arXiv:1909.07528}, 2019.

\bibitem{ma2024efficient}
C.~Ma, A.~Li, Y.~Du, H.~Dong, and Y.~Yang, ``Efficient and scalable
  reinforcement learning for large-scale network control,'' \emph{Nature
  Machine Intelligence}, vol.~6, no.~9, pp. 1006--1020, 2024.

\bibitem{da2019survey}
F.~L. Da~Silva and A.~H.~R. Costa, ``A survey on transfer learning for
  multiagent reinforcement learning systems,'' \emph{Journal of Artificial
  Intelligence Research}, vol.~64, pp. 645--703, 2019.

\bibitem{amatya2022shall}
S.~Amatya, M.~Ghimire, Y.~Ren, Z.~Xu, and W.~Zhang, ``When shall i estimate
  your intent? costs and benefits of intent inference in multi-agent
  interactions,'' in \emph{2022 American Control Conference (ACC)}.\hskip 1em
  plus 0.5em minus 0.4em\relax IEEE, 2022, pp. 586--592.

\bibitem{sinapov2015learning}
J.~Sinapov, S.~Narvekar, M.~Leonetti, and P.~Stone, ``Learning inter-task
  transferability in the absence of target task samples,'' in \emph{Proceedings
  of the 2015 international conference on autonomous agents and multiagent
  systems}, 2015, pp. 725--733.

\bibitem{braylan2016object}
A.~Braylan and R.~Miikkulainen, ``Object-model transfer in the general video
  game domain,'' in \emph{Proceedings of the AAAI Conference on Artificial
  Intelligence and Interactive Digital Entertainment}, vol.~12, no.~1, 2016,
  pp. 136--142.

\bibitem{barrett2015cooperating}
S.~Barrett and P.~Stone, ``Cooperating with unknown teammates in complex
  domains: A robot soccer case study of ad hoc teamwork,'' in \emph{Proceedings
  of the AAAI Conference on Artificial Intelligence}, vol.~29, no.~1, 2015.

\bibitem{littman2001friend}
M.~L. Littman \emph{et~al.}, ``Friend-or-foe q-learning in general-sum games,''
  in \emph{Proceedings of the International Conference on Machine Learning},
  vol.~1, 2001, pp. 322--328.

\bibitem{barreto2017successor}
A.~Barreto, W.~Dabney, R.~Munos, J.~J. Hunt, T.~Schaul, H.~P. van Hasselt, and
  D.~Silver, ``Successor features for transfer in reinforcement learning,''
  \emph{Advances in neural information processing systems}, vol.~30, 2017.

\bibitem{liu2022efficient}
W.~Liu, L.~Dong, D.~Niu, and C.~Sun, ``Efficient exploration for multi-agent
  reinforcement learning via transferable successor features,'' \emph{IEEE/CAA
  Journal of Automatica Sinica}, vol.~9, no.~9, pp. 1673--1686, 2022.

\bibitem{hinrichs2011transfer}
T.~Hinrichs and K.~D. Forbus, ``Transfer learning through analogy in games,''
  \emph{Ai Magazine}, vol.~32, no.~1, pp. 70--70, 2011.

\bibitem{sharma2007transfer}
M.~Sharma, M.~P. Holmes, J.~C. Santamar{\'\i}a, A.~Irani, C.~L. Isbell~Jr, and
  A.~Ram, ``Transfer learning in real-time strategy games using hybrid
  cbr/rl.'' in \emph{IJCAI}, vol.~7, 2007, pp. 1041--1046.

\bibitem{cheng2017multi}
C.~Cheng, Z.~Zhu, B.~Xin, and C.~Chen, ``A multi-agent reinforcement learning
  algorithm based on stackelberg game,'' in \emph{2017 6th Data Driven Control
  and Learning Systems (DDCLS)}.\hskip 1em plus 0.5em minus 0.4em\relax IEEE,
  2017, pp. 727--732.

\bibitem{liu2019value}
Y.~Liu, Y.~Hu, Y.~Gao, Y.~Chen, and C.~Fan, ``Value function transfer for deep
  multi-agent reinforcement learning based on n-step returns.'' in
  \emph{IJCAI}.\hskip 1em plus 0.5em minus 0.4em\relax Macao, 2019, pp.
  457--463.

\bibitem{chen2021shall}
Y.~Chen, L.~Zhang, T.~Merry, S.~Amatya, W.~Zhang, and Y.~Ren, ``When shall i be
  empathetic? the utility of empathetic parameter estimation in multi-agent
  interactions,'' in \emph{2021 IEEE International Conference on Robotics and
  Automation (ICRA)}.\hskip 1em plus 0.5em minus 0.4em\relax IEEE, 2021, pp.
  2761--2767.

\bibitem{hu2014accelerating}
Y.~Hu, Y.~Gao, and B.~An, ``Accelerating multiagent reinforcement learning by
  equilibrium transfer,'' \emph{IEEE transactions on cybernetics}, vol.~45,
  no.~7, pp. 1289--1302, 2014.

\bibitem{barreto2020fast}
A.~Barreto, S.~Hou, D.~Borsa, D.~Silver, and D.~Precup, ``Fast reinforcement
  learning with generalized policy updates,'' \emph{Proceedings of the National
  Academy of Sciences}, vol. 117, no.~48, pp. 30\,079--30\,087, 2020.

\bibitem{gimelfarb2021risk}
M.~Gimelfarb, A.~Barreto, S.~Sanner, and C.-G. Lee, ``Risk-aware transfer in
  reinforcement learning using successor features,'' \emph{Advances in Neural
  Information Processing Systems}, vol.~34, pp. 17\,298--17\,310, 2021.

\bibitem{abdolshah2021new}
M.~Abdolshah, H.~Le, T.~K. George, S.~Gupta, S.~Rana, and S.~Venkatesh, ``A new
  representation of successor features for transfer across dissimilar
  environments,'' in \emph{International Conference on Machine Learning}.\hskip
  1em plus 0.5em minus 0.4em\relax PMLR, 2021, pp. 1--9.

\bibitem{szepesvari1996generalized}
C.~Szepesv{\'a}ri and M.~L. Littman, ``Generalized markov decision processes:
  Dynamic-programming and reinforcement-learning algorithms,'' in
  \emph{Proceedings of International Conference of Machine Learning}, vol.~96,
  1996.

\bibitem{littman1994markov}
M.~L. Littman, ``Markov games as a framework for multi-agent reinforcement
  learning,'' in \emph{Machine learning proceedings 1994}.\hskip 1em plus 0.5em
  minus 0.4em\relax Elsevier, 1994, pp. 157--163.

\bibitem{bertsekas2012dynamic}
D.~Bertsekas, \emph{Dynamic programming and optimal control: Volume I}.\hskip
  1em plus 0.5em minus 0.4em\relax Athena scientific, 2012, vol.~4.

\bibitem{zhu2020online}
Y.~Zhu and D.~Zhao, ``Online minimax q network learning for two-player zero-sum
  markov games,'' \emph{IEEE Transactions on Neural Networks and Learning
  Systems}, vol.~33, no.~3, pp. 1228--1241, 2020.

\bibitem{howard1960dynamic}
R.~A. Howard, ``Dynamic programming and markov processes.'' 1960.

\bibitem{bertsekas1996neuro}
D.~Bertsekas and J.~N. Tsitsiklis, \emph{Neuro-dynamic programming}.\hskip 1em
  plus 0.5em minus 0.4em\relax Athena Scientific, 1996.

\bibitem{lee2023finite}
D.~Lee, ``Finite-time analysis of minimax q-learning for two-player zero-sum
  markov games: Switching system approach,'' \emph{arXiv preprint
  arXiv:2306.05700}, 2023.

\bibitem{fernandez2010probabilistic}
F.~Fern{\'a}ndez, J.~Garc{\'\i}a, and M.~Veloso, ``Probabilistic policy reuse
  for inter-task transfer learning,'' \emph{Robotics and Autonomous Systems},
  vol.~58, no.~7, pp. 866--871, 2010.

\bibitem{anderson2018evaluation}
P.~Anderson, A.~Chang, D.~S. Chaplot, A.~Dosovitskiy, S.~Gupta, V.~Koltun,
  J.~Kosecka, J.~Malik, R.~Mottaghi, M.~Savva \emph{et~al.}, ``On evaluation of
  embodied navigation agents,'' \emph{arXiv preprint arXiv:1807.06757}, 2018.

\bibitem{lehnert2017advantages}
L.~Lehnert, S.~Tellex, and M.~L. Littman, ``Advantages and limitations of using
  successor features for transfer in reinforcement learning,'' \emph{arXiv
  preprint arXiv:1708.00102}, 2017.

\bibitem{lowe2017multi}
R.~Lowe, Y.~I. Wu, A.~Tamar, J.~Harb, O.~Pieter~Abbeel, and I.~Mordatch,
  ``Multi-agent actor-critic for mixed cooperative-competitive environments,''
  \emph{Advances in neural information processing systems}, vol.~30, 2017.

\bibitem{yu2020meta}
T.~Yu, D.~Quillen, Z.~He, R.~Julian, K.~Hausman, C.~Finn, and S.~Levine,
  ``Meta-world: A benchmark and evaluation for multi-task and meta
  reinforcement learning,'' in \emph{Conference on robot learning}.\hskip 1em
  plus 0.5em minus 0.4em\relax PMLR, 2020, pp. 1094--1100.

\end{thebibliography}
\bibliographystyle{IEEEtran}
\end{document}


\section{Proof of Theoretical Results}
\subsection{Proof of Proposition I}

\textbf{Theorem 1. (Game Generalized Policy Improvement)} Let $\Pi_1, \Pi_2, ... \Pi_n $ be $n$ decision policies and let $\tilde Q ^{\Pi_1}, \tilde Q^{\Pi_2},... \tilde Q ^{\Pi_n}$ be approximates of their respective action value function such that
\begin{align*}
|Q ^{\Pi_i}(s,a, b)- \tilde Q ^{\Pi_i}(s,a, b)| \leq \epsilon \\
\text{ for all } s \in S , a \in A, b \in B \text{ and } i \in \{1,2,... n\}.
\end{align*}

Define
\begin{align*}
\pi(s) \in arg\max_a \min_b \min_i \tilde Q^{\Pi_i} (s,a, b)
\end{align*}

We start with the assumptions necessary for this learning algorithm to satisfy the conditions of Theorem 1 in \cite{littman1996generalized} and therefore converge to optimal Q values. The dynamic programming operator defining the optimal Q function is.
Proof. Simplifying the notation, let\\
\begin{align*}
Q_{min}(s,a,b)= min_{i} Q^{\Pi_i}(s,a,b) \text{ and } \\
\tilde{Q}_{min}(s,a,b) = \min_i \tilde{Q}^{\Pi_i}(s,a,b)
\end{align*}
We start by noting that for any $s \in S$ and any $a \in A$ and any $b \in B$ the following holds:
\begin{align*}
 |Q_{min}(s,a, b) - \tilde Q_{min}(s,a,b)|=  |\min_i  Q^{\Pi_i}(s,a, b) - \\
 \min_i  \tilde Q^{\Pi_i}(s,a,b)|=  \min_i| Q^{\Pi_i}(s,a, b) - \tilde Q^{\Pi_i}(s,a,b)| \leq \epsilon
\end{align*}

This property should remain at a minimum as well as a maximum.

For all $s \in S \text{ and } a \in A \text{ and } i \in {1, 2}$, we have
\begin{align*}
T^\Pi\tilde{Q}_{min}(s,a, b)= r(s,a, b) + \sum_s p (s'|s, a, b) \tilde{Q}_{min}(s', \Pi(s')) \\
= r(s,a,b) + \sum_s p (s'|s, a, b) \max_a \min_b \tilde{Q}_{min}(s', a, b)  \\
\geq   r(s,a,b) + \sum_s p (s'|s, a, b) \max_a \min_b {Q}_{min}(s', a, b) - \gamma \epsilon  \\
\text{ this is property of Bellman Operator} \\
\geq   r(s,a,b) + \sum_s p (s'|s, a,b){Q}_{min}(s', \Pi_i(s')) - \gamma \epsilon  \\
\geq   r(s,a,b) + \sum_s p (s'|s, a,b){Q}^{\Pi_i}(s', \pi_i(s')) - \gamma \epsilon  \\
=T^{\Pi_i} Q^{\Pi_i} (s,a,b) - \gamma \epsilon \\
=Q^{\Pi_i}(s,a,b) -\gamma \epsilon
\end{align*}

Since $T^\Pi \tilde Q_{min} (s,a, b) \geq Q^\Pi_i(s,a, b) -\gamma \epsilon$ for any $i$ task, it must be the case that
\begin{align*}
T^\Pi \tilde Q _{min} (s,a, b) \geq \min_i Q^{\Pi_i}(s,a) -\gamma \epsilon\\
= Q_{min}(s,a) - \gamma \epsilon\\
\geq \tilde Q_{min} -\epsilon -\gamma \epsilon
\end{align*}

The Bellman operator in reinforcement learning is said to have two key properties: monotonicity and contraction.

\textbf{Monotonicity}:

Definition: A mapping $T$ is said to be monotonic if, for any two functions $
V_1$ and $V_2$ such that $V_1 \leq V_2$, pointwise, it follows that 
$ TV_1 \leq TV_2$ pointwise. \\
In the context of the Bellman operator: If  $Q_1 \leq Q_2$ pointwise (meaning
 $Q_1 (s,a) \leq Q_2 (s,a)$ for all s and a), then it implies that $ TQ_1 \leq TQ_2$. In other words, improving the estimate of the Q-values for state-action pairs will result in an improved estimate after applying the Bellman operator.
 
\textbf{Contraction (or contraction mapping) property}:

Definition: A mapping $T$ is a contraction if there exists a constant $
0\leq \gamma  < 1$ such that, for all functions $V_1$ and $V_2$, it follows that $|| TV_1 - TV_2|| \leq \gamma ||V_1- V_2||$ where$||.||$ denotes some norm.

In the context of the Bellman operator: If $T$ is a contraction, applying the Bellman operator to two different Q-value functions results in Q-value functions that are closer together. This property is particularly useful in iterative algorithms because it guarantees convergence to a unique fixed point.

In summary, the monotonicity property ensures that improvements in the Q-value estimates lead to improvements after applying the Bellman operator, and the contraction property guarantees the convergence of iterative methods to a unique solution.

These properties are crucial in the analysis of reinforcement learning algorithms, especially those based on iterative methods like value iteration or Q-learning. They provide theoretical guarantees on the convergence of the algorithms and the consistency of the estimated values.

Now we look into the fixed point theorem:
Simplifying the Bellman operator under the assumptions of a deterministic policy and a constant function $e(s,a)=1$ for all s,a.

Starting with the Bellman operator:
\begin{equation}
T^\pi Q(s,a) = \sum_{s'} P(s'|s,a)[R(s,a,s')+\gamma \sum_{a'} \pi(a'|s')Q(s',a')].
\end{equation}
Assumptions:
$\pi(a'|s')=1$ (deterministic policy) \\
$e(s,a)=1$ for all s,a

Now, let's apply these assumptions to simplify the Bellman operator:
\begin{align*}
T^\pi (\tilde Q_{min} (s,a) + ce(s,a)) = \sum_{s'} P(s'|s,a)[R(s,a,s')+\\
\gamma \sum_{a'} \pi (a'|s')( \tilde Q_{min}(s',a')+ce)].
\end{align*}
Given the deterministic policy, $\pi(a'|s')=1$, so the summation over a' simplifies:
\begin{align}
&= \sum_{s'} P(s'|s,a)[R(s,a,s')+\gamma ( \tilde Q_{min}(s',a')+ce)].
\end{align}
\text{Now, since $e(s,a)=1$ for all s,a, the term $ce(s,a)$ becomes}
\text{c, and we have:}
\begin{align}
&=  \sum_{s'} P(s'|s,a)[R(s,a,s')+\gamma ( \tilde Q_{min}(s',a')+c)].
\end{align}
Expanding the sum over s' and using the fact that $P(s'|s,a)$ is a probability distribution and considering the fact that $\pi(a'|s') = 1$ implies that there is only one relevant a' for each s', we can simplify further:
\begin{align}
&= R(s,a,s')+\gamma(\tilde Q_{min}(s',a',)+c)]
\end{align}
\text{Finally, the expression becomes:}\\
\begin{align}
&= \tilde Q_{min}(s,a)+ \gamma c
\end{align}

Now, we look into how the property can be used for multi-agent settings, if the policy is deterministic the above property holds:

Now we want to prove that:

First and foremost:

\begin{align*}
 Q^\Pi(s,a, b)  = \lim_{k \rightarrow \infty}(T^\pi)^k  \tilde Q_{min}(s,a, b) \\
 = lim_{k+1 \rightarrow \infty}(T^\pi)^{k+1} T ^\pi \tilde Q_{min}(s,a, b) \\
  \geq lim_{k+1 \rightarrow \infty}(T^\pi)^{k} (\tilde Q_{min}(s,a, b) - \epsilon (1+ \gamma)) \\
    = lim_{k+2 \rightarrow \infty}(T^\pi)^{k} T^\pi (\tilde Q_{min}(s,a, b) - \epsilon (1+ \gamma)) \\
    = lim_{k+2 \rightarrow \infty}(T^\pi)^{k} (\tilde Q_{min}(s,a, b) - \gamma \epsilon (1+ \gamma))\\
 \geq \tilde Q_{min}-\epsilon \frac{1+ \gamma}{1- \gamma} \text{ : geometric expansion if $\gamma \leq 1$}\\ 
= Q_{min}(s,a, b) - \epsilon -\epsilon \frac{1+ \gamma}{1- \gamma}
\end{align*}

 Proof: the result is a direct application of the theorem 1 and Lemma 1. For any $j$

\subsection{Proof of Lemma 1}
\textbf{Lemma 1.}: Let $\delta_{ij} = \max_{s, a, b} \lvert r_i(s, a, b) - r_j(s, a, b) \rvert$ and let $\Pi$ be an arbitrary policy. 
Then,
\[
\lvert Q^{\Pi}_i(s, a, b) - Q^{\Pi}_j(s, a, b) \rvert \leq \frac{\delta_{ij}} {1 - \gamma} 
\]

\textbf{Proof.} 
Let us simplify the notation, now let $Q^j_i(s,a,b) = Q^{\Pi_j^*}_i(s,a,b).$ Then,
\begin{align*}
 Q^i_i(s,a,b) - Q^j_i(s,a,b)=  Q^i_i(s,a,b)-  Q^j_j(s,a,b)+  \\
 Q^j_j(s,a,b)-  Q^j_i(s,a,b)
 \leq \\\lvert Q^i_i(s,a,b)-  Q^j_j(s,a,b) \rvert+ \lvert Q^j_j(s,a,b)-  Q^j_i(s,a,b)\rvert
\end{align*}
Our strategy will be to bound  $\lvert Q^i_i(s,a,b)-  Q^j_j(s,a,b) \rvert$ and $\lvert Q^j_j(s,a,b)-  Q^j_i(s,a,b)\rvert$ Note that $\lvert Q^i_i(s,a,b)-  Q^j_j(s,a,b) \rvert$ is the difference between the value functions of two Markov Games with the same transition function but potentially different rewards.

Define $\Delta_{ij} = \max_{s, a, b} \lvert Q^{i}_i(s, a, b) - Q^{j}_j(s, a, b) \rvert$. Then,
\begin{align*}
&\lvert Q^{i}_i(s, a, b) - Q^{j}_j(s, a, b) \rvert \\
&= \lvert r_i(s, a, b) + \sum_{s'} p(s' \mid s, a, b) \max_{a' \in A} \min_{b' \in B} Q^{\Pi}_i(s', a', b') \\
&\quad - r_j(s, a, b) - \sum_{s'} p(s' \mid s, a, b) \max_{a' \in A} \min_{b' \in B} Q^{\Pi}_j(s', a', b') \rvert \\
&= \lvert r_i(s, a, b) - r_j(s, a, b) + \sum_{s'} p(s' \mid s, a, b) \big( \max_{a' \in A} \min_{b' \in B} Q^{\Pi}_i(s', a', b') \\
&\quad - \max_{a' \in A} \min_{b' \in B} Q^{\Pi}_j(s', a', b') \big) \rvert \\
&\leq \lvert r_i(s, a, b) - r_j(s, a, b) \rvert + \sum_{s'} p(s' \mid s, a, b) \lvert \max_{a' \in A} \min_{b' \in B} Q^{\Pi}_i(s', a', b') \\
&\quad - \max_{a' \in A} \min_{b' \in B} Q^{\Pi}_j(s', a', b') \rvert \\
&\leq \delta_{ij} + \Delta_{ij}.
\end{align*}


We now turn our attention to $\lvert Q^j_j(s,a,b)-  Q^j_i(s,a,b)\rvert$. Following the previous step: define $\Delta_{ij}' = \max_{s, a, b} \lvert Q^{j}_j(s, a, b) - Q^{j}_i(s, a, b) \rvert$. Then,

\begin{align*}
&\lvert Q^{j}_j(s, a, b) - Q^{j}_i(s, a, b) \rvert \\
&= \left\lvert r_j(s, a, b) + \gamma \sum_{s'} p(s' \mid s, a, b) Q^{j}_j(s', \Pi_j^*(s')) \right. \\
&\quad \left. - r_i(s, a, b) - \gamma \sum_{s'} p(s' \mid s, a, b) Q^{j}_i(s', \Pi_j^*(s')) \right\rvert \\
&= \left\lvert r_i(s, a, b) - r_j(s, a, b) \right. \\
&\quad \left. + \gamma \sum_{s'} p(s' \mid s, a, b) \left( Q^{j}_j(s', \Pi_j^*(s')) - Q^{j}_i(s', \Pi_j^*(s')) \right) \right\rvert \\
&\leq \lvert r_i(s, a, b) - r_j(s, a, b) \rvert \\
&\quad + \gamma \sum_{s'} p(s' \mid s, a, b) \lvert Q^{j}_j(s', \Pi_j^*(s')) - Q^{j}_i(s', \Pi_j^*(s')) \rvert \\
&\leq \delta_{ij} + \Delta_{ij}'.
\end{align*}

Solving for the $\Delta_{ij}'$, you get\\
\begin{equation}
\Delta_{ij}' \leq \frac{\delta_{ij}} {1 - \gamma} 
\end{equation}

Now for the desired result:
\begin{align*}
     Q^i_i(s,a,b) - Q^j_i(s,a,b) \leq   \lvert Q^i_i(s,a,b)-  Q^j_j(s,a,b) \rvert+ \\
     \lvert Q^j_j(s,a,b)-  Q^j_i(s,a,b)\rvert \leq \frac{2\delta_{ij}} {1 - \gamma} 
\end{align*}


\textbf{Proposition 1}.
Let $M_i \in \mathcal{M}^\phi$ and let $Q_i^{\Pi_j^*}$ be the value function of an optimal policy of $M_j \in \mathcal{M^\phi}$ when executed in $M_i$. Given the set $\{ \tilde Q_i^{\Pi^*_1},  \tilde Q_i^{\Pi^*_2}, ...  \tilde Q_i^{\Pi^*_n}\}$ such that
\begin{align*}
\lvert Q^{\Pi_j^*}_i(s, a, b) - \tilde Q^{\Pi_j^*}_i(s, a, b) \rvert \leq \epsilon \textit{ for all } s \in S, \\
a\in A \textit{ and } j \in {1,2,... n},
\end{align*}
let,\\
\begin{align*}
\pi(s) \in arg\max_a \min_b \min_j \tilde Q^{\Pi_j}_i (s,a, b)
\end{align*}
Then,\\
\begin{align*}
Q^*_i(s,a,b)- Q^\Pi_i(s,a) \\
\leq \frac{2}{1-\gamma} \max_{s,a,b} \lvert r_i(s,a,b) -r_j (s,a,b) \rvert  + \frac{2}{1-\gamma} \epsilon 
\end{align*}

Proof, The result is a direct application of Theorem 1 and Lemmas 1 and 2. For any $j \in \{1, 2, ... n\}$, we have.
\begin{align*}
Q^*_i(s,a,b)- Q^\Pi_i(s,a)
\\\leq Q^*_i(s,a,b)- Q^j_i(s,a) + \frac{2}{1-\gamma} \epsilon \text{ Theorem 1 }\\
\leq  \frac{2}{1-\gamma} \delta_{ij} + \frac{2}{1-\gamma} \epsilon \text{ Lemma 1 }\\
= \frac{2}{1-\gamma} \max_{s,a,b} \lvert r_i(s,a,b) -r_j (s,a,b) \rvert  + \frac{2}{1-\gamma} \epsilon 
\end{align*}

\section{Implementation Details}
In this section we describe in detail of the environmental setup and training details of our empirical studies. Pursuer Evader is a standard experiment for zero sum game, we adopted the hyperparameters used in the \cite{barreto2017successor}. We also introduced a challenging task in the Pursuer Evader game which has different initial conditions and more possible goals.


\begin{figure}[b]
    \centering
    \includegraphics[width=1\linewidth]{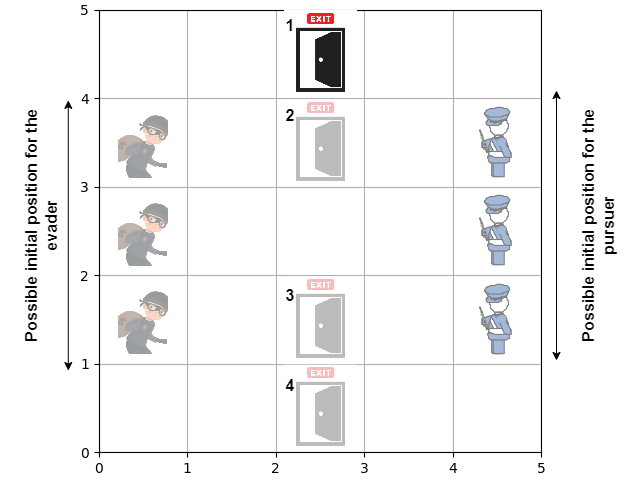}
        \caption{All the initial position and goal for the quantitative test.}
    \label{fig:summary_quan}
\end{figure}
\subsection{Pursuer-Evader for Reward and Policy Transfer}
Section 5 of the paper provides an intuitive overview of the pursuer-evader game used in our experiments. In our setup, when transitioning to a new task, the successor features are initialized using those learned from the previous task. Specifically, at the start of Task 2, the successor feature table from Task 1 is copied, and the policy is initialized using the GGPI algorithm. We use a discount factor of $\lambda = 0.9$, which is a standard choice and yielded the best results in our tests.

\subsubsection{Preliminary Tests}
Two key features in the pursuer-evader tasks are: (1) the distance between the agents, $d(x_e - x_o)$, and (2) the distance between the evader and potential goal positions, $d(x_e - g_n)$, where $n$ denotes the set of possible goals/tasks. In our preliminary experiments, we conducted task transfer from Task 1 to Task 2 using 20,000 training iterations on Task 1, without incorporating terminal rewards. However, this setup did not lead to improvements in cumulative returns or effective policy transfer.

To address this, we introduced a terminal reward feature, $r_t(g_n)$, and additionally incorporated feature weights $w$ to balance the influence of the two distance-based features. An ablation study revealed that the weights listed in Table 1 yielded the best results in terms of both reward transfer and one-shot policy transfer in Task 2.

\subsection{Pursuer Evader Qualitative Test}
 In this section we provide more information on the pursuer evader game and possible output for both agents with different initial positions. As seen in Fig: \ref{fig:summary_quan}, there are total of 9 combination of initial position for the agents to move towards the goal and the the exit number infront of each door is the tasks.

Because of the increase in the number of the doors, the new task is described by weights, the length of weights differs from 8 to 10. Hence , like in Case study 1, the task is given by. $[0.7,-1.3, 0.7, 0, 0, 0, 0, 0,0 ]$ where the first weight is for manhattan distance between the agents, the next two parameter for the Task 1 ( the weight for distance from evader to goal and the weight of terminal reward for evader), after that the subsequent two are for the next task in the increasing order.

\subsection{Algorithms}
\begin{figure}[h]
    \centering
    \includegraphics[width=1\linewidth]{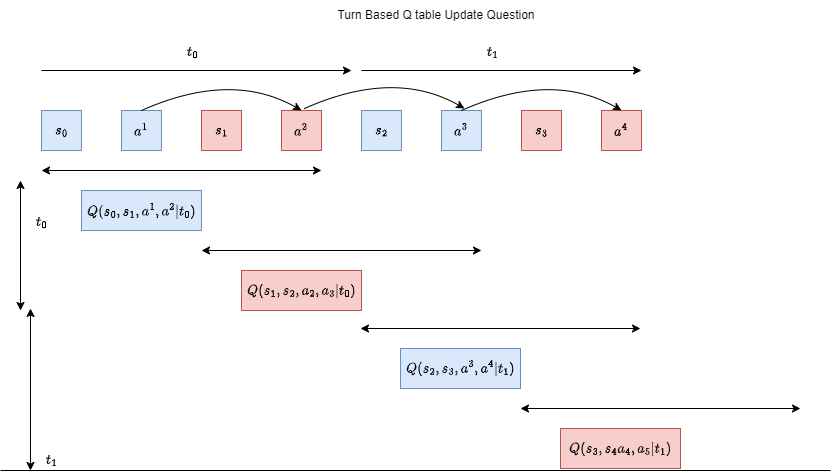}
        \caption{Asynchronous q table update with one step look out horizon for both agents.}
    \label{fig:lookout_horizon}
\end{figure}

As mentioned in the Section 4 of the paper, we have both agents updating their TD error at the same time. Figure \ref{fig:lookout_horizon} shows a time scale of how both agents use a one step horizon look out for the other agent's action and choose to maximize based on the current reward. 

This has been implemented for all the algorithms where both agents are able to have a one step look ahead into the opponent's policy.

\begin{figure*}[t]
    \centering
    \includegraphics[width=0.8\linewidth]{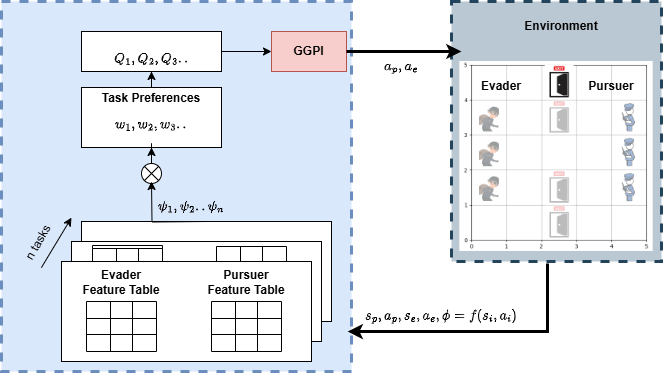}
    \caption{Schematic representation of how the $\psi$ is taken from the table and GGPI is implemented. Given a state $s \in S$, the ﬁrst step is to compute,for each $a*b \in A \times B$, the values of the n SFs: $\psi^{\pi_i} (s, a,b)$. 
    We have sf-table for each task for both pursuer and evader. Based on the agent who is taking the turn, we choose the table. After this, we calculate n Q values by the multiplication of the SFs with Task Preferences $w$, the result of which is an $n \times |A| \times |B| $matrix whose $(i, a,b)$ entry is $Q^{\pi_i} (s, a,b )$. GPI then consists in ﬁnding the minimum element of this matrix and returning the associated actions $a$ and $b$.}
    \label{fig:arcitecture_test}
\end{figure*}

Fig. 3 provides a schematic view of the computation of $\psi$ and the use GGPI to find actions at a timestep.

\bibliography{ref}
\bibliographystyle{IEEEtran}